\documentstyle[12pt,fleqn,epsf,epsfig]{article}

\begin{document}

\title{From Coupled Dynamical Systems to Biological Irreversibility}

\author{Kunihiko Kaneko\\
{\small \sl Department of Pure and Applied Sciences,
College of Arts and Sciences,}\\
{\small \sl University of Tokyo,}\\
{\small \sl Komaba, Meguro-ku, Tokyo 153, Japan}\\
}

\date{}

\maketitle
\begin{abstract}

In the first half of the paper, some recent advances in
coupled dynamical systems, in particular, a globally coupled map are surveyed.
First, dominance of Milnor attractors in 
partially ordered phase is demonstrated.  Second,  chaotic itinerancy in 
high-dimensional dynamical systems
is briefly reviewed, with discussion on a
possible connection with a Milnor attractor network.  
Third, infinite-dimensional collective dynamics is studied, in the
thermodynamic limit of the globally coupled map, where
bifurcation to lower-dimensional attractors by the addition of noise is
briefly reviewed.

Following the study of coupled dynamical systems, 
a scenario for developmental process of cell society is proposed, based on
numerical studies of a system with interacting units with internal dynamics and
reproduction.  Differentiation of cell types is
found as a natural consequence of such a system.
``Stem cells" that either proliferate or differentiate to different types 
generally appear in the system, where irreversible
loss of multipotency is demonstrated.
Robustness of the developmental process against microscopic and macroscopic
perturbations is found and explained, while
irreversibility in developmental process is analyzed in terms of the gain of stability,
loss of diversity and chaotic instability.  Construction of a
phenomenology theory for development is discussed in comparison with the
thermodynamics.

\end{abstract}

\section{Introduction}

How should we understand the origin of biological irreversibility?

As an empirical fact, we know that the direction from the alive to the dead is
irreversible.  At a more specific level, we know that,  
in a multicellular-organism with a developmental process, there is
a definite temporal flow; Through the developmental process, the multipotency,
i.e., the ability to create a different type of cells, decreases.
Initially, the embryonic stem cell has totipotency, and has the potentiality to
create all types of cells in the organism.  Then a stem cell can create 
a limited variety of cells, having multipotency.  This hierarchical loss of
multipotency terminates at a determined cell, which can only replicate its
own type, in the normal developmental process.
The degree of determination increases in the normal course of
development.  How can one understand such irreversibility?

Of course this question is not easy to answer.  However, it
should be pointed that

1) It is very difficult to imagine that this 
irreversibility is caused by a set of specific genes.
The present irreversibility is too universal to be attributed to characteristics of a few molecules.

2) It is also impossible to simply attribute this irreversibility
to the second law of thermodynamics.  One can hardly imagine that
the entropy, even if it were possible to be defined,
suddenly increases at the death, or successively increases at the cell differentiation process.
Furthermore, it should be generally very difficult to define a 
thermodynamic entropy to a highly nonequilibrium system such as a cell.

Then what strategy should we choose?

A biological system contains always sufficient degrees of freedom,
say, a set of chemical concentrations in a cell, which change in time.
Then, one promising strategy for the study of a biological system
lies in  the use of dynamical systems\cite{Turing}.  By setting a class of dynamical
systems, we search for universal characteristics that are robust against microscopic and macroscopic fluctuations.

A biological unit, such as a cell, has
always some internal structure that can change in time.
As a simple representation, the unit can be represented by a dynamical system. For example,
consider a representation of a cell by a set of chemical concentrations.
A cell, however, is not separated from the outside world completely.
For example, isolation by a biomembrane is flexible and incomplete.
In this way, the units, represented by dynamical systems,
interact with each other through the external environment.
Hence, we need a model consisting of
the interplay between inter-unit and intra-unit dynamics.
For example, the complex chemical reaction dynamics
in each unit (cell) is affected by the interaction with other cells,
which provides an interesting example of
``intra-inter dynamics".
In the  `intra-inter dynamics',
elements having internal dynamics interact with each other.
This type of intra-inter dynamics is not necessarily represented only by
the perturbation of the internal dynamics by the interaction with
other units, nor is it merely a
perturbation of the interaction by adding some internal dynamics.

As a specific example of the scheme of intra-inter dynamics, we will mainly
discuss the developmental process of a cell society accompanied by
cell differentiation.
Here, the intra-inter dynamics consists of several biochemical reaction
processes.  The cells interact through the diffusion of
chemicals or their active signal transmission.  

If $N$ cells with $k$ degrees of freedom exist, the total
dynamics is represented by an $Nk$-dimensional dynamical
system (in addition to the degrees of freedom of the environment).
Furthermore, the number of cells is not
fixed in time, but they are born by division (and die) in time.

After the division  of a cell,
if two cells remained identical,
another set of variables would not be necessary.
If the dynamical system for chemical state of a cell 
has orbital instability (such as chaos), however,
the orbits of chemical dynamics of the (two) daughters will diverge.
Hence, the number of degrees of freedom, $Nk$, changes in time.
This increase in the number of variables is tightly connected with the
internal dynamics.
It should also be noted that in the developmental process, in general,
the initial condition of the cell states
is chosen so that their reproduction continues.
Thus, a suitable initial condition for the internal degrees of freedom
is selected through interaction.

Now, to study a biological system in terms of dynamical systems theory, 
it is first necessary to understand the behavior of
a system with internal degrees of freedom and interaction\cite{KK-Tsuda},
This is the main reason why I started a model called Coupled Map Lattice \cite{CML}
(and later  Globally  Coupled Map\cite{GCM}) about 18 years ago.
Indeed, several discoveries in GCM seem to be relevant to understand some basic features
in a biological system.  
GCM has provided us some novel concepts for non-trivial dynamics between
microscopic and macroscopic levels, while the dynamic complementarity between
a part and the whole is important to study biological organization\cite{Complexity}.
In the present paper, we briefly review the behaviors
of GCM in \S 2, and discuss some recent advances in \S 3 - \S 5,
about dominance of Milnor attractors, chaotic itinerancy, and collective  dynamics.  
Then we will switch to the topic of development and differentiation in an interacting cell 
system.  After presenting our model based on dynamical systems in \S 6, we
give a basic scenario discovered in the model, and interpret cell differentiation in terms 
of dynamical systems.  Then, the origin of biological irreversibility is
discussed in \S 9.  Discussion towards the construction of phenomenology theory of
development is given in \S 10.

\section{High-dimensional chaos revisited}

The simplest case of global interaction is studied
as the ``globally coupled map" (GCM) of chaotic elements \cite{GCM}.
A standard  example is given by

\begin{equation}
x_{n+1}(i)=(1-\epsilon )f(x_{n}(i))+\frac{\epsilon }{N}\sum_{j=1}^N f(x_{n}(j))
\end{equation}
where $n$ is a discrete time step and $i$ is the index of an
element ($i= 1,2, \cdots ,N$ = system size), and $f(x)=1-ax^{2}$.
The model is just a mean-field-theory-type extension of coupled map
lattices (CML)\cite{CML}.  

Through the interaction, elements are tended to oscillate synchronously,
while chaotic instability leads to destruction of the coherence.
When the former tendency wins, all elements oscillate coherently,
while elements are completely desynchronized in the limit of
strong chaotic instability.  Between these cases, elements split
into clusters in which they oscillate coherently.
Here a cluster is defined
as a set of elements in which $x(i)=x(j)$ \cite{GCM}.
Attractors in GCM are classified by the number of synchronized
clusters $k$ and the number of elements for each cluster $N_i $.
Each attractor is coded by the
clustering condition $[k,(N_1 ,N_2 ,\cdots,N_k )]$.
Stability of each clustered state is analyzed by
introducing the split exponent \cite{GCM,KK94}.

An interesting possibility in the clustering is
that it provides a source for diversity.
In clustering it should be noted that identical chaotic elements differentiate
spontaneously into different groups:
Even if a system consists of identical elements,
they split into groups with different phases of oscillations.
Hence a network of chaotic elements gives
a theoretical basis for isologous diversification
and provides a mechanism for the
origin of diversity and complexity in biological networks \cite{Alife,Rel}.

In a globally coupled chaotic system in general,
the following phases appear successively with the increase of
nonlinearity in the system ($a$ in the above logistic map case)
\cite{GCM}:

(i) {\bf Coherent phase}: Only a coherent attractor ($k=1$) exists.

(ii) {\bf Ordered phase}: All attractors consist of few ($k=o(N)$) clusters.

(iii) {\bf Partially ordered phase}: Attractors with a variety of clusterings
coexist, while most of them have many clusters ($k=O(N)$).

(iv) {\bf Turbulent phase}:  Elements are completely desynchronized, and
all attractors have $N$ clusters.

The above clustering behaviors
have universally been confirmed in a variety of systems.

In the partially ordered (PO) phase,
there are a variety of attractors with a different number of clusters,
and a different way of partitions $[N_1,N_2, \cdots ,N_k]$.
The clustering here is typically inhomogeneous:  The partition
$[N_1,N_2, \cdots ,N_k]$ is far from equal partition.  
Often  this clustering is hierarchical as for the number of elements, and
as for the values.  For example, consider the following idealized clustering:
First split the system into two equal clusters.  Take one of them and split it again
into two equal clusters, while leave the other without split.
By repeating this process, the partition is given by
$[N/2,N/4,N/8,\cdots]$.   In this case, the difference of the values of $x_n(i)$
is also hierarchical.  The difference between the values of $x_n(i)$ decreases as 
the above process of partition is iterated.
Although the above partition is too much simplified, such hierarchical structure
in partition and in the phase space is typically observed
in the PO phase. The partition is organized as an inhomogeneous tree structure, as in
the spin glass model \cite{SG}.

We have also measured the fluctuation of the partitions,
using the probability $Y$ that two elements fall on the same cluster.
In the PO phase, this $Y$ value fluctuates by initial conditions, and
the fluctuation remains finite even if the size goes to infinity \cite{GCM-part,Vul}.
It is noted that such remnant fluctuation of partitions is also seen in
spin glass models \cite{SG}.

\section{Chaotic Itinerancy and Milnor attractors}

In the Partially ordered (PO) phase, there coexist
a variety of attractors depending on the partition\cite{GCM-part}.
To study the stability of an attractor against perturbation, we introduce the
return probability $P(\sigma)$, defined as follows\cite{GCM-Milnor}:
Take an orbit point $\{x(i)\}$ of an attractor in an
$N$-dimensional phase space,
and perturb the point to $x(i)+\frac{\sigma}{2}rnd_i$, where
$rnd_i$ is a random number taken from $[-1,1]$, uncorrelated for all elements
$i$. Check if this perturbed point returns to the original attractor
via the original deterministic dynamics (1).
By sampling over random perturbations and the time of the application of
perturbation, the return probability $P(\sigma)$ is defined as
(\# of returns )$/$ (\# of perturbation trials).
As a simple index for robustness of an attractor, it is
useful to define $\sigma_c$ as the largest $\sigma$ such that $P(\sigma)=1$.
This index measures what we call the {\sl strength} of an attractor.

The strength $\sigma_c$ gives a minimum distance between
the orbit of an attractor and its basin boundary.
In contrast with our naive expectation from the concept of
an attractor, we have often observed
`attractors' with $\sigma_c =0$, i.e.,
$P(+0) \equiv \lim_{\delta \rightarrow 0}P(\delta) <1$.
If $\sigma_c = 0$ holds for a given state, it cannot be an
``attractor" in the sense with asymptotic stability,
since some tiny perturbations kick the orbit out of the ``attractor".
The attractors with $\sigma_c =0$
are called Milnor attractors\cite{Milnor0,Milnor}.  In other words,
Milnor attractor is defined as an attractor that is unstable
by some perturbations of arbitrarily small size, but globally attracts
orbital points.  The basin of attraction has a positive Lebesgue measure.
(The basin is riddled here \cite{riddle,riddle-CM}.)
Since it is not asymptotically
stable, one might, at first sight, think that it is
rather special, and appears only at a critical point like the
crisis in the logistic map\cite{Milnor0}.  
To our surprise, 
the Milnor attractors are rather commonly observed around the PO phase
in our GCM.  The strength and basin volume of attractors are not necessarily correlated.
Attractors with $\sigma _c=0$ often have a large basin volume.

Still, one might suspect that such Milnor attractors must be weak
against noise.  Indeed, by a very weak noise with the amplitude $\sigma$,
an orbit at a Milnor attractor is kicked away, and if the orbit is reached 
to one of attractors with $\sigma_c >\sigma$, it never comes back to the Milnor attractor.
Rather, an orbit kicked out from a Milnor attractor is often found to stay in the vicinity of
it\cite{Milnor}. The orbit comes back to the original Milnor attractor
before it is kicked away to other attractors with $\sigma_c >\sigma$.
Furthermore, by a larger noise, orbits sometimes are more attracted 
to Milnor attractors.  Such attraction is possible, since Milnor attractors
here have global attraction in the phase space, in spite of their local
instability.

Dominance of Milnor attractors gives us to suspect the computability of
our system.  Once the digits of two variable $x(i)=x(j)$ 
agree down to the lowest bit, the values never split again,
even though the state with the synchronization of the two elements may be
unstable.  As long as digital computation is adopted,
it is always possible that an orbit is trapped to such unstable state.
In this sense a serious problem is cast in numerical computation
of GCM in general \footnotemark.   

\footnotetext{Indeed, in our simulations 
we have often added a random floating at the
smallest bit of $x(i)$ in the the computer, to 
partially avoid such computational problem.}

Existence of Milnor attractors may lead us to suspect the
correspondence between a (robust) attractor and memory,
often adopted in neuroscience (and theoretical cell biology).
It should be mentioned that Milnor attractors can provide
dynamic memory \cite{Tsuda,KK-Tsuda} allowing for 
interface between outside and inside, external inputs and internal
representation.

\section{Chaotic Itinerancy}

Besides the above {\sl static} complexity, {\sl dynamic} complexity
is more interesting at the PO phase.  Here orbits make itinerancy over
ordered states with partial synchronization of elements, via highly chaotic states.
This dynamics, called chaotic itinerancy (CI),
is a novel universal class in high-dimensional dynamical systems.
Our CI consists of a quasi-stationary
high-dimensional state, exits to ``attractor-ruins" with low effective degrees
of freedom, residence therein, and chaotic exits from them.
In the CI, an orbit successively itinerates over
such ``attractor-ruins", ordered motion with some coherence among elements.
The motion at ``attractor-ruins" is quasistationary.
For example, if the effective degrees of freedom is two,
the elements split into two groups, in each of which elements
oscillate almost coherently.  The system is in the vicinity of a
two-clustered state, which, however, is not a stable attractor, but keeps 
attraction to its vicinity globally within the phase space.
After staying at an attractor-ruin, an orbit exits from it
due to chaotic instability, and shows a high-dimensional chaotic
motion without clear coherence.  This high-dimensional state
is again quasistationary, although there are some holes connecting
to the attractor-ruins from it.  Once the orbit is trapped at a hole, it
is suddenly attracted  to one of attractor ruins, i.e., ordered states with low-dimensional
dynamics.

This CI dynamics
has independently been found in a model of neural dynamics by Tsuda \cite{Tsuda},
optical turbulence \cite{Ikeda}, and in GCM.
It provides an example
of successive changes of relationships among elements.

Note that the Milnor attractors satisfy
the condition of the above ordered states constituting chaotic itinerancy.
Some Milnor attractors we have found keep
global attraction, which is consistent with the observation
that the attraction to ordered states in chaotic itinerancy
occurs globally from a high-dimensional chaotic state.
Attraction of an orbit to precisely a given attractor requires infinite time, and before 
the orbit is really settled to a given
Milnor attractor, it may be kicked away\footnotemark.
When Milnor attractors that lose the stability ($P(0)<1$) keep global attraction,
the total dynamics can be constructed as
the successive alternations to the attraction to, and
escapes from, them.  
If the attraction to robust attractors from a given Milnor attractor is
not possible, the long-term dynamics with the noise strength $\rightarrow +0$ is
represented by successive transitions over Milnor attractors.
Then the dynamics is represented by transition matrix over among Milnor attractors.
This matrix is generally asymmetric: often, there is a
connection from a Milnor attractor A to a Milnor attractor B, but
not from B to A.  The total dynamics is represented by the motion over
a network, given by a set of directed graphs over Milnor attractors.

\footnotetext{This problem is subtle computationally, since any finite
precision in computation may have a serious influence on whether the orbit remains 
at a Milnor attractor or not.}

In general, the `ordered states' in CI may not be exactly Milnor attractors but can be
weakly destabilized states from Milnor attractors.  Still,
the attribution of CI to Milnor attractor network dynamics is expected to
work as one ideal limit \footnotemark.

\footnotetext{The notion of chaotic itinerancy is rather broad,
and some of CI may not be explained by the Milnor attractor
network.  In particular, chaotic itinerancy in
a Hamiltonian system\cite{Konishi,Shinjo}
may not fit directly with the present correspondence.}

As already discussed about the Milnor attractor,
computability of the switching over Milnor attractor networks has a serious problem.
In each event of switching, which Milnor attractor is visited next
after the departure from a Milnor attractor may depend on the precision.
In this sense, the order of visits to Milnor attractors in chaotic itinerancy
may not be undecidable in a digital computer.  In other words, motion at a macroscopic
level may not be decidable from a microscopic level.
With this respect, it may be interesting to note
that there are similar statistical features between (Milnor attractor)
dynamics with a riddled basin
and undecidable dynamics of a universal Turing-machine\cite{Saito-KK}.

\section{Collective Dynamics}

If the coupling strength~$\epsilon$ is small enough,
oscillation of each element has no mutual synchronization.
In this turbulent phase, $x(i)$ takes almost random values almost independently, and 
the number of degrees of freedom is proportional to the number of
elements $N$, i,e., the Lyapunov dimension increases in proportion to $N$.
there remains some coherence among elements.
Even in such case, the macroscopic motion shows some coherent motion
distinguishable from noise, and there remains some coherence among elements, 
even in the limit of $N \rightarrow \infty$.
As a macroscopic variable we adopt the mean field,
\begin{equation}
h_n={1\over N}\sum_{i=1}^Nf(x_n(i)).
\end{equation}
In almost all the parameter values,
the mean field motion shows some dynamics that is distinguishable from noise,
ranging from torus-like to higher dimensional motion.
This motion remains even in  the thermodynamic limit\cite{KK-col}.

This remnant variation means that the collective dynamics $h_n(i)$ keeps
some structure.  One possibility is that the dynamics is
low-dimensional.  Indeed in some system with a global coupling,
the collective motion is shown to be low-dimensional
in the limit of $N \rightarrow \infty$ ( see \cite{Kurths,Shibata97}.)
In the GCM eq.(1), with the logistic or tent map, low-dimensional
motion is not detected generally, although there remains some collective motion
in the limit of $N \rightarrow \infty$.
The mean field motion in GCM is regarded to be infinite
dimensional, even when the torus-like motion is
observed~\cite{Ershov,Chawanya,Nakagawa,Shibata98}.
Then it is important to clarify the nature of this mean-field dynamics.

It is not so easy to examine the infinite dimensional dynamics, directly. Instead,
Shibata, Chawanya and the author have first made the motion low-dimensional
by adding noise, and then studied the limit of noise $\rightarrow 0$.
To study this effect of noise, we have simulated the model
\begin{equation}
x_{n+1}(i)=(1-\epsilon )f(x_n (i))+\frac{\epsilon }{N}\sum_{j=1}^N f(x_n (j))
+\sigma \eta_n ^i ,
\end{equation}
where $\eta_n ^i$ is a white noise generated by an
uncorrelated random number homogeneously distributed over [-1,1].

The addition of noise can destroy the above coherence among elements.
In fact, the microscopic external noise leads
the variance of the mean field distribution to
decrease with $N$~\cite{KK-col,Perez0}.
This result also implies decrease of the mean field fluctuation by external noise.

Behavior of the above equation
in the thermodynamic limit~$N\rightarrow\infty$ is represented by
the evolution of the one-body distribution function~$\rho_n(x)$ at time step~$n$ directly.
Since the mean field value
\begin{equation}
h_n=\int f(x)\rho_n(x)dx
\label{eq:meanfiled}
\end{equation}
is independent of each element,
the evolution of $\rho_n(x)$ obeys the Perron-Frobenius
equation given by,
\begin{equation}
\rho_{n+1}(x)=\int dy{1\over\sqrt{2\pi}\sigma}
e^{-{\left(F_{n}(y)-x\right)^2\over2\sigma^2}}
\rho_{n}(y),
\label{eq:PF}
\end{equation}
with
\begin{equation}
F_{n}(x)=(1-\epsilon)f(x)+\epsilon h_{n}.
\end{equation}

By analyzing the above Perron-Frobenius equation\cite{Shibata99}, it is shown that
the dimension of the collective motion increases as $log (1/\sigma ^2)$,
with $\sigma$ as the noise strength.  Hence in the limit of $\sigma \rightarrow 0$,
the dimension of the mean field motion is expected to be infinite.
Note that the mean field dynamics~(at~$N\rightarrow\infty$) is
completely deterministic, even under the external noise.

With the addition of noise, high-dimensional structures in the mean-field
dynamics are destroyed successively, and the bifurcation from
high-dimensional to low-dimensional chaos, and then to
torus proceeds with the increase of the
noise amplitude.  With a further increase of noise to $\sigma > \sigma _c$,
the mean field goes to a fixed point through Hopf bifurcation.
This destruction of the hidden coherence leads to a strange
conclusion.  Take a globally coupled system with a
desynchronized and highly chaotic state, and add  noise to the system.
Then the dimension of the mean field motion gets lower with the increase of noise.

The appearance of low-dimensional `order' through the
destruction of small-scale structure in chaos is
also found in noise-induced order\cite{Tsuda-Matsumoto}.  Note however that 
in a conventional noise-induced transition\cite{NIP},
the ordered motion is still stochastic, since the noise is added into a low-dimensional dynamical
system.  On the other hand, the noise-induced transition in the collective dynamics occurs
after the thermodynamic limit is taken.  Hence the low-dimensional dynamics induced
by noise is truly low-dimensional.  When we say a torus, the Poincare map shows a curve 
without thickness
by the noise, since the thermodynamic limit smears out the fluctuation around the tours.
Also, it is interesting to note that a similar  mechanism of the
destruction of hidden coherence is observed in  quantum chaos.

This noise-induced low-dimensional collective dynamics
can be used to distinguish high-dimensional chaos from random noise.
If the irregular behavior is originated in  random noise,
 (further) addition of noise will result in an increase of
the fluctuations.  If the external application of noise leads to
the decrease of fluctuations in some experiment, it is natural to
assume that the irregular dynamics there is due to high-dimensional chaos
with a global coupling of many nonlinear modes or elements.

\section{Cell Differentiation and development as dynamical systems}

Now we come back to the problem of cell differentiation and development.
A cell is separated from environment by a membrane,
whose separation, however, is not complete.
Some chemicals pass through the membrane, and
through this transport, cells interact with each other.
When a cell is represented by a dynamical system
the cells interact with each other and with the external environment.
Hence, we need a model consisting of the interplay between inter-unit and intra-unit dynamics.
Here we will mainly
discuss the developmental process of a cell society accompanied by
cell differentiation, where
the intra-inter dynamics consist of several biochemical reaction
processes.  Cells interact through the diffusion of
chemicals or their active signal transmission, while
they divide into two when some condition is satisfied with
the chemical reaction process in it. (See Fig.1 for schematic representation of our model).

We have studied several models\cite{KY94,KY97,KY99,KK97,FK98a,FK98b} 
with (a) internal (chemical) dynamics of several degrees of freedom,
(b) cell-cell interaction type through the medium, and
(c) the division to change the number of cells.

As for the internal dynamics, auto-catalytic reaction among
chemicals is chosen. Such auto-catalytic
reactions are necessary to produce chemicals in a
cell, required for reproduction\cite{Eigen}.  
Auto-catalytic reactions often lead to nonlinear oscillation in chemicals.
Here we assume the possibility of such oscillation in the intra-cellular dynamics
\cite{Goodwin,Hess}.
As the interaction mechanism, the diffusion
of chemicals between a cell and its surroundings is chosen.

To be specific, we mainly consider the following model here.
First, the state of a cell $i$  is assumed to be characterized by the cell volume and
a set of functions $x^{(m)} _i(t)$ representing the concentrations of $k$ chemicals
denoted by $m=1,\cdots,k$.  The concentrations of chemicals change as a result of
internal biochemical reaction dynamics within each cell and
cell-cell interactions communicated through the surrounding medium.

For the internal chemical reaction dynamics, we choose a catalytic network among
the $k$ chemicals.  The network is defined by a collection of triplets ($\ell$,$j$,$m$)
representing the reaction from chemical $m$ to $\ell$  catalyzed by $j$.
The rate of increase of $x^{\ell}_i(t)$ (and decrease of $x^{m}_i(t)$)
through this reaction is given by $x^{(m)}_i(t) (x^{(j)}_i(t))^{\alpha}$, where
$\alpha$ is the degree of catalyzation
($\alpha = 2$ in the simulations considered presently).
Each chemical has several paths to other chemicals, and thus a complex
reaction network is formed.  The change in the chemical concentrations through all such
reactions, thus, is determined by the set of all terms of
the above type for a given network.
(These reactions can include genetic processes).

Cells interact with each other through the transport of chemicals out of and into
the surrounding medium.  As a minimal case, we consider only indirect
cell-cell interactions through diffusion of chemicals via the medium.
The transport rate of chemicals into a cell is proportional to
the difference in chemical concentrations between the inside and the outside of
the cell, and is given by $D(X^{(\ell)} (t) -x^{({\ell})}_i (t))$,
where $D$ denotes the diffusion constant, and $X^{(\ell)} (t)$ is the concentration of
the chemical at the medium. 
The diffusion of a chemical species through cell membrane should depend on
the properties of this species.
In this model, we consider the simple case in which there are two types of
chemicals, one that can penetrate the membrane and one that cannot.
For simplicity, we assume that all the chemicals capable of penetrating the membrane
have the same diffusion coefficient, $D$.
With this type of interaction, corresponding chemicals in the medium are consumed.
To maintain the growth of the organism, the system is immersed
in a bath of chemicals through which (nutritive) chemicals are supplied to the cells.

As chemicals flow out of and into the environment,
the cell volume changes.  The volume  is assumed to be proportional 
to the sum of the quantities of chemicals in the cell, and thus is a dynamical variable.
Accordingly, chemicals are diluted as a result of the increase of the cell volume.

In general, a cell divides according to its internal state, for example,
as some products, such as DNA or the membrane, are
synthesized, accompanied by an increase in cell volume.  Again,
considering only a simple situation, we assume that a cell divides into two
when the cell volume becomes double the original.
At each division,  all chemicals are almost equally divided,
with random fluctuations.

Of course, each result of simulation depends on the specific choice of the reaction network.
However, the basic feature of the process to be discussed
does not depend on the details of the choice, as long as the
network allows for the oscillatory intra-cellular dynamics leading to the
growth in the number of cells.
Note that the network is not constructed to imitate
an existing biochemical network.
Rather, we try to demonstrate that important features in a biological system
are a natural consequence of a system with internal dynamics,
interaction, and reproduction.  From the study we try to extract a
universal logic underlying a class of biological systems.

\section{Scenario for Cell Differentiation}

From several simulations of the models
starting from a single cell initial condition, we have shown that
cells undergo spontaneous differentiation as the number is increased.
(see Fig.2 for schematic representation):
The first differentiation starts with the clustering of the phase of the oscillations,
as discussed in globally coupled maps (see Fig.2a).  Then, the differentiation comes to the stage
that the average concentrations of the biochemicals over the cell
cycle become different.  The composition of
biochemicals as well as the rates of catalytic reactions and transport of
the biochemicals become different for each group.

After the formation of cell types, the chemical compositions
of each group are inherited by their daughter cells. In other words,
chemical compositions of cells are recursive over divisions.
The biochemical properties of a cell are inherited by its progeny, or in other
words, the properties of the differentiated cells are stable, fixed or
determined over the generations (see Fig. 2b).  
After several divisions, such initial condition of units is
chosen  to give the next generation of the same  type as its mother cell.

The most interesting example here is the formation of
stem cells, schematically shown given in Fig.2c \cite{FK98a}. This cell type, denoted as `S' here, 
either reproduces the same type or forms different cell types,
denoted for example as type A and type B.  Then after division
events $S \rightarrow S,A,B$ occur.  Depending on the adopted chemical
networks, the types A and B replicate, or switch to different 
types.  For example $A \rightarrow A, A1,A2,A3$ is observed in
some network.  This hierarchical organization is
often observed when the internal dynamics have some complexity,
such as chaos.

The differentiation here is ``stochastic", arising from chaotic 
intra-cellular chemical dynamics.
The choice for a stem cell either to replicate or to differentiate
looks like stochastic as far as the cell type is
concerned.  Since such stochasticity
is not due to external fluctuation but is a result of the internal state,
the probability of differentiation can be regulated by the intra-cellular state.
This stochastic branching is accompanied by
a regulative mechanism.  When some cells are removed externally during 
the developmental process, the rate of differentiation changes so that
the final cell distribution is recovered.

In some biological systems such
as the hematopoietic system, stem cells  either replicate or
differentiate into different cell type(s).  This differentiation
rule is often hierarchical \cite{Alberts,Ogawa}.
The probability of differentiation to one of the several blood cell
types is expected to depend on the interaction.
Otherwise, it is hard to explain why the developmental
process is robust.  For example, when the number of some terminal cells
decreases, there should be some mechanism to increase the rate of
differentiation from the stem cell to the differentiated cells.
This suggests the existence of interaction-dependent regulation of the differentiation 
ratio, as demonstrated in our results.

{\sl Microscopic Stability}

The developmental process is stable against molecular fluctuations.
First, intra-cellular dynamics of each cell type are stable against such
perturbations.  Then, one might think that this selection of each cell type is 
nothing more than a choice among basins of attraction
for a multiple attractor system.  If the interaction were neglected,
a different type of dynamics would be interpreted as a different attractor.
In our case, this is not true, and cell-cell interactions are necessary to 
stabilize cell types.  Given cell-to-cell interactions,
the cell state is stable against perturbations on the level of
each intra-cellular dynamics. 

Next, the number
distribution of cell types is stable against fluctuations.  Indeed,
we have carried out simulations of our model, by adding a noise term,
considering finiteness in the number of molecules\cite{KY99,FK01}.
The obtained cell type as well as the number distribution is hardly affected
by the noise as long as the noise amplitude is not too large.\footnote{
When the noise amplitude is too large, distinct types are no longer formed.
Cell types are continuously distributed.  In this case, the division speed 
is highly reduced,  since the differentiation of roles by differentiated 
cell types is destroyed.}

{\sl Macroscopic Stability}

Each cellular state is also stable against perturbations of the interaction
term.  If the cell type number distribution is changed within some range, 
each cellular dynamics keeps its type.  Hence discrete, stable
types are formed through the interplay between intra-cellular dynamics
and interaction. 
The recursive production is attained through the selection of initial
conditions of the intra-cellular dynamics of each cell, so that it is rather
robust against the change of interaction terms as well.

The macroscopic stability is clearly shown in the 
spontaneous regulation of differentiation ratio.
How is this interaction-dependent rule formed?
Note that depending on the distribution of the other cell types,
the orbit of internal cell state is slightly deformed.
For a stem cell case,
the rate of the differentiation or the replication
(e.g., the rate to select an arrow among $S\rightarrow S,A,B$)
depends on the cell-type distribution.  For example,
when the number of ``A" type cells is reduced, the orbit of an ``S-"type cell
is shifted towards the orbits of ``A", with which the
rate of switch to ``A" is enhanced.  
The information of the cell-type distribution is represented by
the internal dynamics of ``S"-type cells, and it is essential to the regulation of 
differentiation rate \cite{FK98a}.

It should be stressed that our dynamical differentiation
process is always accompanied by this kind of regulation process, without any
sophisticated programs implemented in advance.
This autonomous robustness provides a novel viewpoint to 
the stability of the cell society in multicellular organisms.

\section{Dynamical Systems Representations of Cell Differentiation}

Since each cell state is realized as a balance between internal dynamics
and interaction, one can discuss which part is more relevant to determine
the stability of each state.
In one limiting case, the state is an attractor as internal dynamics\cite{Kauffman},
which is sufficiently stable and not destabilized by cell-cell interaction.
In this case, the cell state is called `determined', according to the terminology in cell biology.
In the other limiting case, the state is totally governed by the
interaction, and by changing the states of other cells, the cell state in concern
is destabilized.  In this case, each cell state is highly dependent on the
environment or other cells.

Each cell type in our simulation generally lies between these two limiting cases.
To see such intra-inter nature of the determination explicitly,
one effective method is a transplantation experiment.
Numerically, such experiment is carried out
by choosing determined cells (obtained from the normal differentiation process)
and putting them into a different set of surrounding cells,
to set distribution of cells so that it does not appear through the normal course
of development.

When a differentiated and recursive cell is transplanted to another cell society,
the offspring of the cell keep the same type,
unless the cell-type distribution of the society is strongly biased.
When a cell is transplanted into a biased society, differentiation from
a `determined' cell occurs.  For example, a homogeneous society consisting only of one 
determined cell type is unstable, and some cells start to switch
to a different type.  Hence,
the cell memory is preserved mainly in each individual cell,
but suitable inter-cellular interactions are also necessary to keep it.

Since each differentiated state is not attractor, but
is stabilized through the interaction, we propose to define  
{\sl partial attractor}, to address attraction
restricted to the internal cellular dynamics.
Tentative definition of this partial attractor is as follows;

(1) [internal stability]
Once the cell-cell interaction is specified (i.e., the dynamics of other cells),
the state is an attractor of the internal dynamics.
In other words, it is an attractor when the dynamics is restricted only to the
variable of a given cell.

(2) [interaction stability] The state is stable against change of interaction term, up to
some finite degree.
With the change of the interaction term of the order $\epsilon$, the change in the dynamics
remains of the order of $O(\epsilon)$.

(3) [self-consistency] 
For some distribution of units of cellular states satisfying (1) and (2), 
the interaction term continues to satisfy the condition (1) and (2).

We tentatively call a state satisfying (1)-(3) as partial attractor. Each determined
cell type we found can be regarded as a partial attractor.
To define the dynamics of stem cell in our model, however, we have to slightly
modify the condition of (2) to a `Milnor-attractor' type.
Here, small perturbation to the interaction term (by the increase of
the cell number) may lead the state to switch to a differentiated state. 
Hence, instead of (2), we set the condition:

(2') For some change of interaction with a finite measure,
some orbits remain to be attracted to the state.

So far we have discussed the stability of a state by fixing the number of cells.
In some case, the condition (3) may not be satisfied when the system is developed
from a single cell following the cell division rule.  As for developmental process,
the condition has to be satisfied for a restricted range of cell distribution
realized by the evolution from a single cell.
Then we need to add the condition:

(4)[Accessibility].  The distribution (3) is satisfied from an initial condition of
a single cell and with the increase of the number of cells.

Cell types with determined differentiation observed in our model is regarded as a state
satisfying (1)(2)(3)(4), while the stem cell type is regarded as
a state satisfying (1)(2')(3)(4).

In fact, as the number is increased, some perturbations to the
interaction term is introduced.  In our model, the stem-cell state satisfies 
(2) up to some number, but with the further increase of number, the condition
(2) is no more satisfied and is
replaced by (2').  Perturbation to the interaction term due to the cell number
increase is sufficient to bring about a switch  from
a given stem-cell dynamics to a differentiated cell.  
Note again that the stem-cell type state with weak stability
has a large basin volume when started from a single cell.

\section{ Towards Biological irreversibility irreducible to thermodynamics}

In the normal development of cells, there is clear irreversibility,
resulting from the successive loss of multipotency.  

In our model simulations, this loss of multipotency occurs irreversibly.
The stem-cell type can differentiate to  other types, while the
determined type that appear later only replicates itself.   
In a real organism, there is a hierarchy in determination, and a stem cell is often 
over a progenitor over only a limited range of cell types.
In other words, the degree of determination is also hierarchical.  In our model,
we have also found such hierarchical structure.  So far, we have found only 
up to the second layer of hierarchy in our model with the number of chemicals $k=20$.

Here, the loss of multipotency
dynamics of a stem-type cell exhibit irregular oscillations
with orbital instability and involve a variety of chemicals.
Stem cells with these complex dynamics have a potential to differentiate into several
distinct cell types.
Generally, the differentiated cells always possess simpler cellular dynamics than the 
stem cells, for example, fixed-point dynamics and regular oscillations.

Although we have not yet succeeded in formulating the irreversible
loss of multipotency in terms of a single fundamental quantity (analogous to
thermodynamic entropy), we have heuristically derived a general law describing the
change of the following quantities in all of our numerical
experiments, using a variety of reaction networks\cite{FK00,FK01}.
As cell differentiation progresses through development,

\begin{itemize}

\item
(I) stability of intra-cellular dynamics increases;

\item
(II) diversity of chemicals in a cell decreases;

\item
(III) temporal variations of chemical concentrations decrease, by realizing less chaotic motion.

\end{itemize}

The degree of (I) could be determined by a minimum change 
in the interaction to switch a cell state, by properly extending the
`attractor strength' in \S 3. Initial undifferentiated cells spontaneously
change their state even without the change of the interaction term, while stem cells
can be switched by tiny change in the interaction term.  The degree of determination  
is roughly measured as the minimum perturbation strength required for a
switch to a different state.

The diversity of chemicals (II) can be measured, for example, by $S=-\sum_{j=1}^k p(j) log p(j)$, with
$p(j)=<\frac{x(j)}{\sum_{m=1}^k x(m)}>$, with $<..>$ as temporal average.  
Loss of multipotency in our model
is accompanied by a decrease in the diversity of chemicals and
is represented by the decrease of this diversity $S$.

The tendency (III) is numerically confirmed by the subspace Kolmorogorv-Sinai (KS) entropy
of the internal dynamics for each cell.  Here, this subspace KS entropy is measured as
a sum of positive Lyapunov exponents, in the tangent space restricted only to the
intracellular dynamics for a given cell.   Again, this exponent decreases through the
development.

\section{Discussion: Towards Phenomenology theory of Developmental Process}

In the present paper, we have first surveyed some of
recent progresses in coupled dynamical systems,
in particular globally coupled maps.  Then we discuss some of our recent studies
on the cell differentiation and development, based on coupled dynamical systems
with some internal degrees of freedom and the potentiality to increase the number
of units(cells).  Stability and irreversibility of the developmental process
are demonstrated by the model, and are discussed in terms of dynamical systems.

Of course, results based on a class of models are
not sufficient to establish a
theory to understand the stability and irreversibility in 
development of multicellular organisms.  We need to unveil the logic
that underlies such models and real development universally.
Although mathematical formulation is not yet established, 
supports are given to the following conjecture. 

{\sl Assume a cell with internal chemical reaction network whose degrees of freedom is 
large enough and which interacts each other through the environment.  
Some chemicals are transported from the environment and converted to other chemicals
within a cell.  Through this process the cell volume increases
and the cell is divided.  The, for some chemical networks, each chemical state of a cell remains
to be a fixed point.  In this case, cells remain identical, where
the competition for chemical resources is higher, and the increase of
the cell number is suppressed.  On the other hand, for some reaction networks,
cells differentiate and the increase in the cell number is not suppressed.  
The differentiation of cell types form a hierarchical rule.  
The initial cell types have large chemical diversity and show
irregular temporal change of chemical concentrations.  As the number of cells
increases and the differentiation progresses, irreversible
loss of multipotency is observed.  This differentiation process is triggered by
instability of some states by cell-cell interaction, while the realized states of
cell types and the number distribution of such cell types are
stable against perturbations, following the
spontaneous regulation of differentiation ratio.}

When we recall the history of physics, the most successful phenomenological theory is 
nothing but thermodynamics.
To construct a phenomenology theory for development, or generally a theory for
biological irreversibility,  comparison with the thermodynamics should be relevant.
Some similarity between the phenomenology of development and thermodynamics
is summarized in Table 1.

As mentioned, both the thermodynamics and the development phenomenology
have stability against perturbations.  Indeed, the
spontaneous regulation in a stem cell system found in our model is a clear demonstration of
stability against perturbations, that is common with the Le Chatelier-Braun  principle.
The irreversibility in thermodynamics is defined by suitably restricting possible
operations, as formulated by adiabatic process.
Similarly, the irreversibility in a multicellular organism has to be suitably
defined by introducing an ideal developmental process.  Note that in some
experiments like clone from somatic cells in animals\cite{clone}, the irreversibility 
in normal development can be reversed.  

The last question that should be addressed here is the search for macroscopic quatities
to characterize each (differentiated) cellular 'state'.
Although thermodynamics is established by cutting the macroscopic out of microscopic levels,
in a cell system, it is not yet sure if such macroscopic quantities can be defined, by
separating a macroscopic state from the microscopic level.  At the present stage,
there is no definite answer.
Here, however, it is interesting to recall recent experiments of tissue engineering.
By changing the concentrations of only three control chemicals, Asashima
and coworkers\cite{Asashima} have
succeeded in constructing all tissues from a Xenopus undifferentiated cells
(animal cap).  Hence there may be some hope that a reduction to a few
variables characterizing macroscopic `states' may be possible.

Construction of phenomenology for development charactering its stability
and irreversibility is still at the stage `waiting for Carnot', but
following our results based on coupled dynamical systems models and
some of recent experiments, 
I hope that such phenomenology theory will be realized in near future.

\vspace{.2in}
Table I:
Comparison with development phenomenology with thermodynamics

\hspace{-.3in}
\begin{tabular}{|c||c|c|} \hline

 &development phenomenology& thermodynamics \\ \hline
Stability & cellular and  ensemble level & macroscopic \\ \hline
stability against perturbation & regulation of differentiation ratio & Le Chatelier Braun\\ \hline
irreversibility & loss of multipotency & second law  \\ \hline
quantification of irreversibility& some pattern of gene expression? & entropy  \\ \hline
cycle & somatic clone cycle? &Carnot cycle  \\ \hline
\end{tabular}

\vspace{.2in}

{\sl acknowledgments}

The author is grateful to T. Yomo , C. Furusawa, T. Shibata for discussions.
The work is partially supported by Grant-in-Aids for Scientific
Research from the Ministry of Education, Science, and Culture of Japan.

\begin{figure}
\noindent
\hspace{-.3in}
\epsfig{file=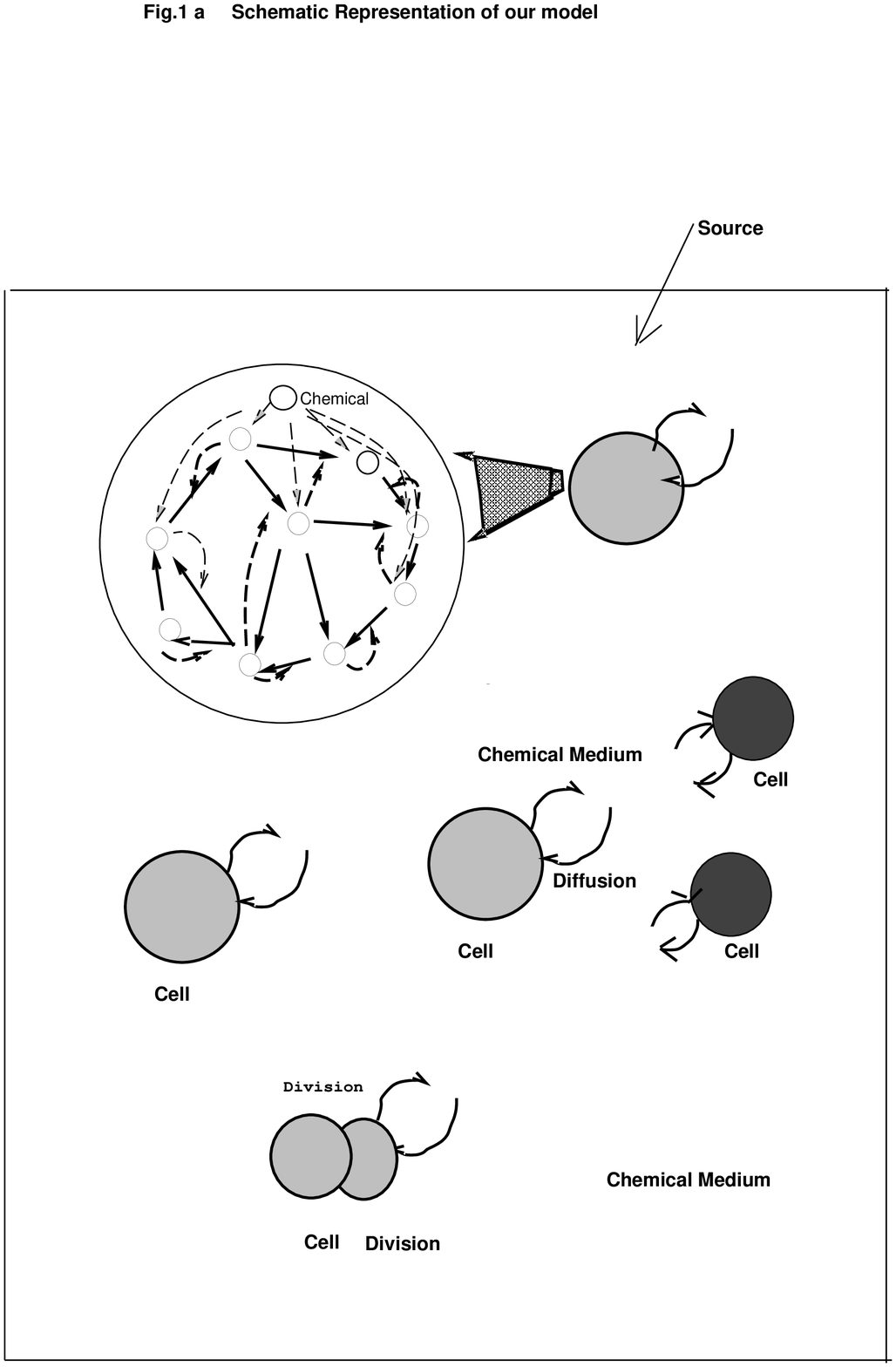,width=.8\textwidth}
\caption{Schematic representation of our model.}
\end{figure}


\hspace{-.3in}
\begin{figure}
\epsfig{file=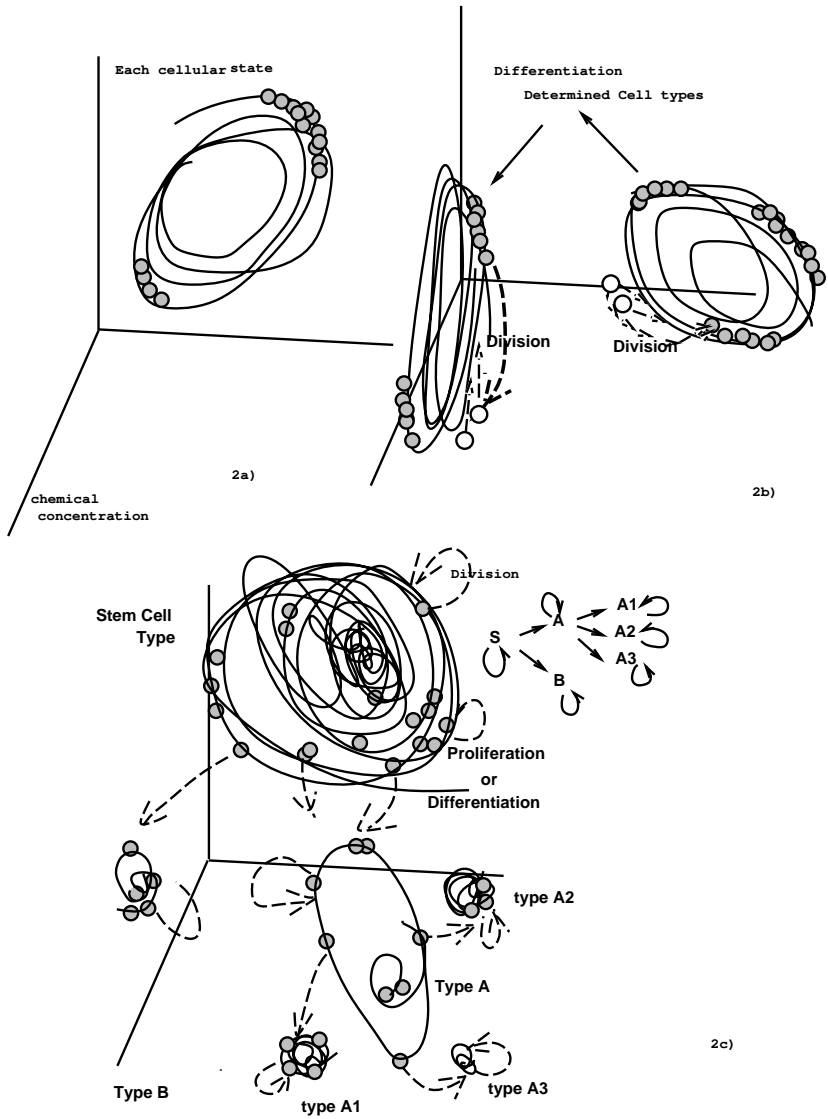,width=.8\textwidth}
\caption{Schematic representation of cell differentiation process, plotted in the phase
space of chemical concentrations.}
\end{figure}

\end{document}